\documentclass[sigconf, screen, authorversion, nonacm]{acmart}

\setcopyright{cc}
\copyrightyear{2025}

\AtBeginDocument{%
  \providecommand\BibTeX{{%
    \normalfont B\kern-0.5em{\scshape i\kern-0.25em b}\kern-0.8em\TeX}}}

\usepackage{wrapfig}

\usepackage{tabularx}

\usepackage{enumitem}

\newlist{questions}{enumerate}{2}
\setlist[questions,1]{label=RQ\arabic*.,ref=RQ\arabic*, itemsep=1pt, topsep=3pt}
\setlist[questions,2]{label=(\alph*),ref=\thequestionsi(\alph*)}

\begin{document}

\title[A Practitioner-Led Investigation of  Public Interest Technologists in Civil Society]{Shaping a Profession, Building a Community: A Practitioner-Led Investigation of  Public Interest Technologists in Civil Society}

\author{Mallory Knodel}
\email{mallory.knodel@nyu.edu}
\affiliation{
  \institution{New York University}
  \institution{Center for Democracy and Technology}
  \city{Washington}
  \state{DC}
  \country{US}}

\author{Mallika Balakrishnan}
\email{mallikajbalakrishnan@gmail.com}
\affiliation{
  \institution{Center for Democracy and Technology}
  \city{London}
  \country{UK}}

\author{Lauren M. Chambers}
\email{lauren@ischool.berkeley.edu}
\affiliation{
  \institution{Berkeley School of Information}
  \institution{Center for Democracy and Technology}
  \city{Berkeley}
  \state{CA}
  \country{US}}

\begin{abstract}
  The label `public interest technology' (PIT) is growing in popularity among those seeking to use `tech for good' - especially among technical practitioners working in civil society and nonprofit organizations. PIT encompasses a broad range of sociotechnical work across professional domains and sectors; however, the trend remains understudied within sociotechnical research. This paper describes a mixed-methods study, designed and conducted by PIT practitioners at the Center for Democracy and Technology, that characterizes technologists within the specific context of civil society, civil rights, and advocacy organizations in North America and Western Europe. We conducted interviews with civil society leaders to investigate how PIT practitioners position the field and themselves, and we held a roundtable discussion bringing diverse voices together to make meaning of this growing phenomenon. Ultimately, we find that PIT remains both defined and plagued by its expansiveness, and that today's civil society public interest technologists see a need for both (a) more robust professionalization infrastructures, including philanthropic attention, and (b) more engaged, coherent community. This study illuminates a nascent intersection of technology and policy on-the-ground that is of growing relevance to critical sociotechnical research on the shifting relationship between computing and society.
\end{abstract}



\received[Uploaded]{10 August 2025}

\maketitle

\section{Introduction}\label{intro}
Critical sociotechnical research offers a long history of studying policymaking around technology \cite{Yang, knot, Spaa, bell_its_2022}, civic and public sector technologies \cite{Johnson, scale, scott_algorithmic_2022}, and the role of technology in advocacy \cite{bikes, bennett, krafft_action-oriented_2021}. Today, these concepts are being explored, refined, and operationalized on the ground by a broad group of practitioners aligning under the label ``public interest technology'' (PIT). Indeed, due to PIT's increasingly central role as a hub for interdisciplinary professionals pioneering technologically-driven solutions to sociotechnical challenges \cite{CSCWworkshop, PowerToThePublic}, it is crucial to develop a more comprehensive grasp of what public interest tech is - and who public interest technologists are - as a specific and growing phenomenon. Such understanding is imperative for mapping the landscape and progress of technological movements towards justice beyond the ivory tower: a key priority of critical sociotechnical researchers.

What do these practitioners mean by \textit{public interest technology}? Most simply, a public interest technologist does exactly what one might expect: work that involves technology, often bringing technological expertise to bear, in the service of the public interest \cite{Chock, PowerToThePublic}. Technology here might mean one or all of: communication technology, information technology, internet technology, or digital technology \cite{Bruce}. The scope of public interest technology is intentionally broad, ranging from building new technology, to helping a team leverage existing digital tools, to leading tech policy discussions with a focus on rights preservation 
\cite{Freedman2013, Freedman2016}. Certainly, much fundamental boundary work remains to be done regarding who are the affected ``publics,'' what constitutes those publics' ``interests,'' and which ``technologies'' are involved \cite{CSCWworkshop}. Depending on how boundaries are drawn, public interest technology may be inclusive of or distinctive from: civic technology, participatory design, tech policy, public science, open source, digital civics, `tech for good', nonprofit IT, public interest cybersecurity, global internet standards, and more \cite{Chock, digitalCivics, Hackler}. (We note that claims or debates concerning where the boundaries of PIT \textit{ought} to be drawn are out of scope for this work.)

What is a \textit{technologist}? This term, too, is intentionally broad: a label that we authors use to self-identify and that we see widely adopted by colleagues and peers (in both casual settings and in formal job titles) as an explicit result of its breadth. A technologist might be a software developer, a data analyst, a systems designer, a digital privacy consultant, or a technology policy lobbyist. Crucially, within civil society models, an individual technologist's work often includes multiple - or sometimes all - such roles as they serve different needs within and across organizations and sectors \cite{chambers_road_nodate}. 
However, while this work chooses as its focus public interest technologists employed by civil society organizations, others 'doing PIT work' may operate in public interest-focused roles within technology companies, in government bodies, or at organizations producing research and journalism about technology.

Despite its ongoing semantic indeterminism, 'public interest technology' has undoubtedly grown in popularity as a label among professionals in the past six years, driven by investment from philanthropic organizations and its adoption by New America's Public Interest Tech University Network (PIT-UN). 
Yet, as the label proliferates and the proto-sector grows, we worry insufficient attention has been paid to the role of so-called public interest technologists within civil society organizations such as our own: the Center for Democracy and Technology (CDT).\footnote{
CDT is based in the United States and in Europe and focuses primarily on shaping technology policy in the public interest.}

As public interest tech practitioners ourselves, we are motivated by a desire to foster intentionality, clarity, and structure around the forward direction of professional public interest technology at 
CDT 
and beyond. This concern pervades all stages of a career that might be labeled as PIT, but is particularly acute for senior technologists in leadership roles. Via a two-phase iterative methodology integrating qualitative interviews and a roundtable discussion with PIT practitioners, this study characterizes the experiences, challenges, and needs of technologists who work within the specific context of civil society, civil rights, and advocacy organizations over the last two decades. While public interest technologists have been doing the work, by one name or another, for years, the continuing growth of the field introduces an urgent need - and opportunity - for technologists to be active in shaping its future. Critically, this paper is both by and for public interest technologists, uniquely taking place within a tech policy-oriented nonprofit and being undertaken by practitioner-researchers. As such, this work maintains an explicit focus towards shaping a future of PIT for ourselves and those who come after.

Our paper contributes to the ongoing discourse surrounding technology, civil society, and societal change in three ways: (1) by bringing PIT professionals explicitly into the fold of academic discussions; (2) by identifying challenges faced by those working at the direct intersection of technology and civil society; and (3) by offering priorities for future research and projects examining and supporting public interest technology within civil society, as determined by PIT practitioners themselves. The paper is organized as follows: we present related literature and studies of PIT-related work in Section \ref{lit}. 
We describe our study methods and limitations in Section \ref{methods}, before presenting our results in Section \ref{results}. Finally, we discuss the implications of our findings in Section \ref{discussion} and offer conclusions and directions for future work in Section \ref{conclusion}.

\section{Background and Related Work}\label{lit}

To set the stage for this research, in this section we review existing scholarship and work pertaining to the intersection of sociotechnical research and civil society, the historical timeline of PIT, and previously identified challenges that PIT faces. We note that the history and analysis of PIT within civil society has not been told within the academic record; rather, its artifacts are found scattered across various organizational websites and reports (and, increasingly, archives of organizational websites and reports). While narrating a comprehensive history of PIT is not the primary aim of this work, we present below the main elements of its history and the main takeaways from relevant organizational analyses in service of contextualizing our own analysis that follows.

\subsection{Sociotechnical Research and Civil Society}

Sociotechnical researchers have long recognized, embraced, and critiqued the motivation of their field towards societal and political change \cite{Kling, Hochheiser}. Yet over the past 15 years, more and more concerning computer-mediated problems have arisen: from social media misinformation to algorithmic surveillance to discriminatory predictive analytics \cite{eubanks_automating_2018, Khadijah, misinfo}. However, the primary focus of sociotechnical research has not necessarily tracked the needs of non-academic actors or practitioners on-the-ground \cite{industry}. Thus the increased focus on the relationship between human computing and public policy in recent years is not surprising, as a primary site where researchers might be able to ``make an impact'' \cite{Lazar10, Lazar15, livingstone_make_2021}. Many scholars go even further, encouraging the retirement of approaches that distinguish technology as separate from policy, instead embracing the inseparable ``knot'' that is tied between policy, technology, and design \cite{Yang, knot, junginger}. Public interest technology exists at the center of this knot, a now well-explored intersection. 

As stated in Section \ref{intro}, a wide collection of fields, disciplines, movements and initiatives may fall within the category of public interest technology. Many of them have been the object of careful and extensive study, including civic technology \cite{saldivar, gordon, boehner, aragon}, tech and data "for good" \cite{green, rider, powell, madianou, aula}; inclusive, participatory, and critical design \cite{designjustice, friedman, shilton, irani, harrington}; tech policy \cite{knot, krafft}; "hacking" and related questions of technological legitimacy and valorization \cite{powell_hacking_2016}; and more. 
Indeed, this paper builds on prior work studying policymaking around technology within research institutions and think tanks \cite{Spaa}, civic technologies and digital public services \cite{Dow, Johnson}, the complex digital practices that advocates adopt in support of civic and political goals \cite{bikes}, and the professionalization process in computing sub-fields \cite{marsden_how_2018, ding_as_2022, bevan_creating_2005}. We look to critical work, too, which examines the challenges researchers face as they pursue social impact \cite{Balestrini, Mutale} and the ways that ``Public Sector Information Systems'' transcend scale across the national, state, and local levels \cite{scale}. Others have documented the interdisciplinary synergies that arise in the creation, conception, and curation of knowledge in multi-disciplinary collaborations to inform policy and practice \cite{borgman_whos_2012}. 

But the question remains: if so many of PIT's component disciplines have already been comprehensively studied, what remains? Is PIT but old wine in a new bottle?
It is not. 
PIT, with its expansive agenda specifying no more than the "public interest," is a moniker which has been more easily embraced by and integrated into \textit{civil society} than its predecessors (some predecessors include government-centered 'civic tech,' design-oriented 'participatory design,' or ad-hoc 'tech for good' initiatives).
The civil society sector\footnote{We note that the term \textit{civil society} has more than one meaning. Here we use it as a shorthand for the sector of organizations and institutions, rather than as a more general term regarding democratic and civic elements of society.} consists of "the plethora of private, nonprofit, and nongovernmental organizations that have emerged in recent decades in virtually every corner of the world to provide vehicles through which citizens can exercise individual initiative in the private pursuit of public purposes" \cite[p.~60]{salamon}. It encompasses the myriad think tanks, nonprofits, and advocacy groups which, especially in the United States context, have become driving forces behind public education campaigns, legislative lobbying efforts, movement litigation, and even the coordination of political action \cite{leroux, galanter, leachman, bennett}. 
As such, establishing a comprehensive understanding of the relationship between technology and civil society - by using the interdisciplinary frameworks and methods offered by social studies of technology - is critical to understanding processes of sociopolitical change in the 21st century. 

While scholars have considered the effects that insurgent information and digital technologies might have on the operation of civil society organizations writ-large \cite{greenpeace, grover}, very little ink has been spilled exploring the specific role of \textit{technologists within civil society} - that is to say, public interest technologists like ourselves. In their 2018 "\#MoreThanCode" report, Costanza-Chock et al. provided a comprehensive snapshot of practitioners of ``technology for justice and equity'' and "technologists for social justice" (T4SJ)
\cite{Chock}. In 2022, Stapleton et al. held a CSCW workshop which encouraged critical discussions PIT, asking who PIT is \textit{for}, and how researchers might ethically go about pursuing work that centers impacted communities \cite{CSCWworkshop}. Yet the interfaces between PIT and civil society remain largely untouched by sociotechnical researchers today.

\subsection{History of PIT}

Technology initiatives seeking to advance the public interest trace back at least to the beginning of the new millennium. Indeed, the internet itself was developed as a decentralized infrastructure project requiring cooperation, shared governance and public funding \cite{kee_dialectical_2010}. Examples located within civil society range from the founding of the Electronic Frontier Foundation in 1990 and the Center for Democracy \& Technology's hiring of John Morris as the head of the Internet Standards, Technology and Policy Project in 2002 to the technology and social change trainings run by the Association for Progressive Communications through the early 2000s, or the 2000 opening of the Stanford Center for Internet and Society or UC Berkeley's various public interest research centers.\footnote{Such histories are documented on the following organizational web pages: "A History of Protecting Freedom Where Law and Technology Collide" by the  Electronic Frontier Foundation: \url{https://www.eff.org/about/history}; "Prominent Lawyer/Technologist Joins CDT to Lead New Project Promoting the Public Interest in Emerging Internet Technology" by the Center for Democracy and Technology: \mbox{\url{https://web.archive.org/web/20020818013555/http://www.cdt.org/press/010403press.shtml}}; "History" by the Association for Progressive Communications: \url{https://www.apc.org/en/about/history}; "About Us" by the Stanford Center for Internet and Society: \url{https://cyberlaw.stanford.edu/about-us}; and "Public Interest Technology at UC Berkeley" by the Berkeley School of Information: \url{https://www.ischool.berkeley.edu/public-interest}}

Such work has increasingly coalesced around public interest technology as a label since 2016, when the Ford, MacArthur, Knight, Mozilla, and Open Society Foundations established \$18 million in grant funding streams for projects focused on PIT as part of the work of the NetGain Partnership, a digitally-oriented philanthropic collaboration established the year prior.\footnote{See "Five Major Foundations Announce Groundbreaking Plans to Develop Public Interest Technologists," a press release by Open Society Foundations on February 16, 2016: \url{https://www.opensocietyfoundations.org/newsroom/five-major-foundations-announce-groundbreaking-plans-develop-public-interest}} The NetGain member foundations continue to dedicate significant philanthropic support to public interest technology initiatives at many levels. For example, collaborative groups like the ``Civil Rights Privacy \& Technology Table'' founded in 2011 by the Ford Foundation allow for public interest technologists from various civil rights, privacy, and technology organizations to gather to work on shared advocacy and strategy.\footnote{See "Lessons from the Table: Civil rights, technology, and privacy," a 2019 Learning Reflections blog post by Lori McGlinchey for the Ford Foundation: \url{https://www.fordfoundation.org/work/learning/learning-reflections/lessons-from-the-table-civil-rights-technology-and-privacy/} and the "About" page by the Civil Rights Table: \url{{https://www.civilrightstable.org/}}} 
A consulting memo written for the Ford Foundation notes the strategic importance of harnessing technologists' desire to do ``meaningful work" \cite{here-to-there}. Arguably the primary focus of philanthropic efforts in public interest technology since 2016 has been strategizing the creation of a talent pipeline. This pipeline involves academic training and exposure to public interest technology as a field.
Indeed, as of 2024 no fewer than 63 American higher education institutions (including many Ivy League and R1 universities) have established PIT initiatives as members of the Public Interest Technology University Network (PIT-UN), after its 2019 launch with support from New America, the Ford Foundation, and the Hewlett Foundation \cite{higher-ed}.\footnote{See the Public Interest Technology University Network: \url{https://pitcases.org/}}

\subsection{Identified Challenges}\label{challenges}
Various consulting and research reports since 2015 have sought to characterize the challenges that PIT practitioners face. Synthesis of such work reveals four primary problems which we describe here: organizational fit, professional scope, uncommon expertise, and scarce resources.

Costanza-Chock et al. found one major challenge is \textit{ambiguity regarding how public interest technologists fit} within organizations and domains; the \#MoreThanCode report found that many public interest technology practitioners were ``self-taught'' and their work in civil society organizations was catalyzed by a specific instance of organizational need. These practitioners cited the importance of conferences (especially the Allied Media Conference, the Internet Freedom Festival, RightsCon, and Aspiration Tech's Non-Profit Dev Summit), mentorship, and fellowship opportunities as integral gateways to their ability to do public interest technology work in public interest organizations \cite{Chock}. Again, this underscores the multiplicity in the types of work that public interest technologists are doing: in many cases, the profile of the work builds from the practitioner's existing expertise, which could be based in internet technology, computer science, and developer work, or not. For example, an increasing number of advocates with backgrounds in civil rights, workers' rights, voting rights, media studies, grassroots organizing and other areas are leading key fights in technology policy. Roughly 58\% of the practitioners surveyed for the \#MoreThanCode report did not identify as technologists at all, despite their work being centered in technology \cite{Chock}.\footnote{Other notable perceptions of ``public interest technology'' included the idea that public interest technology was specific to the nonprofit and NGO sector, given the focus on advocacy and policy. There were also perceptions that ``public interest technology'' connoted a ``top-down, non-diverse framework'' that was ``predominantly white, male, D.C.-focused, and funder-driven.'' For a more in-depth look at the differences in terminology used to talk about the kinds of work referred to here as public interest technology, see the following taxonomy produced by the \#MoreThanCode Report team: http://bit.ly/PITB-taxonomy} 

Relatedly, Costanza-Chock et al. found a common challenge across organizations to be \textit{struggling to find the right scope} for a technologist's work, leading to several problems. One such problem is expecting technologists to do everything, leading the \#MoreThanCode report to note the importance of not ``expect[ing] the one `technologist' on staff to do everything from general IT support, to design processes, to full stack development \cite{Chock}.'' Conversely, there exists the problem of not adequately utilizing the technologist's full potential because of their siloed place within the organizational structure: for example, a public interest technologist might be limited to providing technical expertise to stakeholders, but not fully be able to bring that expertise to bear on policy decisions that take place in different parts of the organizational leadership. To this end, support and communication between the technologist and senior leadership is also frequently noted as a crucial factor for successful integration of public interest technologists (and public interest technology work) \cite{Muñoz}.

As noted in a 2013 Freedman Consulting report on technologists in civil society and government, public interest institutions have also \textit{struggled to find individuals who have both tech expertise and policy understanding}. Even when they are found, there is a noted ``failure of civil society organizations to best deploy technologists...One nonprofit leader said, `The two problems I see are, one, we don't have enough technologists; two, and when people become technologists they're pretty siloed off.' Another nonprofit technology expert added that, in civil society organizations, `There's a lack of appreciation for what technology can do for your work \cite{Freedman2013}.''' 

Finally, another frequent obstacle faced especially by non-tech-focused organizations is \textit{resource scarcity}. Unlike the challenges of appropriately using a hired technologist, this challenge relates more to the process of a civil society organization deciding whether or not it needs a technologist in the first place. One practitioner interviewed noted that it is difficult ``in resource-constrained organizations, particularly organizations that are not technology organizations, to divert resources from their core mission to technology'' and ``convince them that having a technologist is a force multiplier" \cite{Chock}. Given that most civil rights, civil society, and advocacy organizations operate under conditions of resource scarcity (or at the very least, significant reliance on grant and foundation funding) \cite{bennett}, integrating a public interest technologist might understandably be more difficult unless clearly connected to the organization's mission. Equally, it can be difficult for public interest institutions to offer competitive compensation or perceived career opportunities, compared to the private sector \cite{Freedman2013}.

\subsection{Study Motivation}\label{motivation}

This work was directly inspired by our own organization's experience (re)designing a public interest technologist role, and facing many of the exact challenges described above. In 2020, 
our organization, the Center for Democracy and Technology (CDT)
set out to redesign the role of Chief Technology Officer (CTO). The redesign was motivated by a realization that the organic development of the CTO role, at the forefront of 
CDT's history of
interventions in the technical community
for almost two decades,
needed to be more intentionally set up for regular participation in public policy debates and global standards design. The CTO also needed to provide accessible technical expertise to key stakeholders on critical privacy and security issues, in order to ensure that policy decisions in these areas were informed by technological considerations. 

Our CTO redesign process uncovered questions about how technologists at multiple levels within 
CDT
were integrated into the organization's structure. When this work began, technologists at 
CDT
were grouped primarily into a team of technologists, as opposed to distributed through other teams. However, as findings began to indicate that ``siloing'' technologists was a barrier to effective integration, 
CDT
integrated its technologists across its strategic teams, internalizing a collaborative approach that centers issues rather than technologies. We pursued these changes from the understanding that a more collaborative, interdependent structure could help our organization to better leverage interdisciplinary expertise in the context of our work.

This process sparked our curiosity, and we continued exploring more broadly at the field of public interest technology - a field that 
CDT
has been instrumental at building over the years. We quickly realized that our own challenges and solutions were not unique after all; they were reflected in other organizations. As described in Section \ref{challenges}, ours were problems of fit and scope, and our response was to reorganize our professional roles and to establish conduits for mutual education. We wondered: what \textit{else} might be revealed by putting our proverbial heads together across organizational boundaries that is of interest - and of interest not only to public interest technologists, but also to researchers of technology and society? Ultimately, we developed a qualitative research methodology to learn from the PIT community at large, seeking to address the following profession-wide questions:
\begin{em}
(RQ1) How are public interest technologists being integrated into civil society, civil rights, and advocacy organizations?
(RQ2) In what ways is public interest technology a community of practice; in what ways is it not? How do we strengthen communities of practice within and across organisations and sectors? 
(RQ3) What characteristic successes and challenges do public interest technologists experience in their work?  What challenges are unique to integration across sectors; what are the sector-specific challenges?
\end{em}

\section{Methods}\label{methods}


We adopted a multi-phase iterative methodological approach.
First, a preliminary review of existing relevant work informed the recruitment of in-depth interview participants and design of the interview questions. Second, researcher analysis of the interviews illuminated initial findings, while interviewee expertise helped to fill in gaps within the literature review, ultimately presented in Section \ref{lit}. 
Finally, our findings were distilled and refined through the expert feedback and discussion we gathered in the roundtable. Roundtable participants reviewed the research so far, enriching the findings as presented in Section \ref{results} and informing the recommendations presented in \ref{discussion}. Throughout the process, we prioritized the perspective of the practitioner over that of the researcher in an effort to make good on the promises of participatory methods.

\subsection{Interviews}

We used interviews to explore the themes across PIT history in the context of the real-world experiences of technologists who have ``succeeded'' in integration into public interest organizations. Initial interviewees were leading public interest technologists chosen based on 
CDT's
professional network, and further respondents were collected via snowball sampling from interviewees. While we narrowed our focus to those with a more significant professional tenure (as opposed to early-career professionals), we collected a range of experiences within the field both in terms of professional responsibilities and personal identity, prioritizing that the set of interviewees collectively represent public interest technologists in peer organizations, other organizations, academia, and the private sector. From the identified list of potential interviewees, we conducted 11 interviews between May and June 2021. Three of the 11 participants were women, and three were from non-U.S. contexts (see Table \ref{tab:interviewees}). Participants' expertise spans a range of PIT-related projects, including: designing privacy-enhancing technologies, advising international social movements on how to responsibly employ technology, drafting technical standards and public policies in support of digital rights, operationalizing privacy and cybersecurity from within government agencies, and more.

\begin{table*}[t]
    \caption{Characteristics of all interview participants}
      \label{tab:interviewees}
      \centering

    \begin{tabularx}{\linewidth}[t]{cXccc}
        \toprule
         & Description                                                                                                & Country     & Gender  &PIT Tenure* \\ 
        \midrule
        \textit{P1}           & Lawyer and longtime executive of various Internet policy projects within both civil society and government & USA         & Man     & 31+ yrs \\
        \textit{P2}           & Digital rights and internet policy senior executive (both civil society and government) and researcher                            & USA         & Man     & 25-30 yrs \\
        \textit{P3}           & Digital rights and internet policy researcher, civil society chief technologist and executive                                     & USA         & Man     & 25-30 yrs \\
        \textit{P4}           & Computer engineer, corporate tech policy executive, and leader in internet standards                       & USA         & Woman   & 15-20 yrs \\
        \textit{P5}           & Technologist and leader of cooperative movement supporting tech for collective action                      & Mexico      & Man     &  10-14 yrs\\ 
        \textit{P6}           & Technologist for digital rights nonprofit, consultant, activist                                               & USA         & Man     & 15-20 yrs\\ 
        \textit{P7}           & Researcher of tech, human rights, and security and nonprofit executive                                     & USA         & Woman   & 10-14 yrs \\ \
        \textit{P8}           & Civil society nonprofit technologist, activist, security researcher                                        & France      & Man     & 10-14 yrs\\ 
        \textit{P9}           & Privacy engineer, nonprofit tech executive, internet standards leader                                      & USA         & Man     & 25-30 yrs \\\
        \textit{P10}          & Longtime executive of various Internet and digital rights projects within civil society                    & Netherlands & Man     & 25-30 yrs \\ 
        \textit{P11}          & Media scholar, nonprofit executive, web developer, activist                                                & USA         & Woman   & 15-20 yrs \\ \bottomrule 
         & \multicolumn{4}{p{0.8\linewidth}}{\textit{*Note that tenure does not necessarily correlate with age, as different participants came across PIT-related work at different points within their careers}}
    \end{tabularx}
    \centering
\end{table*}

Before the interviews, participants were provided with a brief literature review cataloging the past few decades of public interest technology work (which would become Section \ref{lit}). All participants agreed to an informed consent form which explained the voluntary and safe nature of the study. Interviews were conducted by the first author via a digital platform; the interviewer later transcribed notes from recorded audio \cite{Zoom}. Virtual interviews allowed us to reach geographically distant participants, and also allowed us to conduct interviews during the COVID-19 pandemic. Interviews were semi-structured and lasted approximately 45-60 minutes. Questions addressed the history, patterns, and challenges of public interest technology as well as personal reflections on being a PIT professional.

Following the interviews, we employed thematic analysis to draw out key takeaways from transcripts relating to the state of PIT within civil society \cite{Weiss, deterding}. The second author coded the interview transcripts and notes using the ATLAS.ti qualitative data analysis software, identifying 15 initial broad themes that emerged.\footnote{Those initial themes included: leadership, archetypes (variety of types of work), legitimacy, pipeline, gaps in the PIT model, diversity, role of various sectors in PIT, collaboration across organisations, common values, lack of community, love of community, multiple community, community outside work, policy community, and the importance of teaching and learning.} Over multiple meetings, and in consultation with other relevant experts as we will describe in the next subsection, the authors compared and discussed key reflections, in the process condensing our list of themes. Ultimately we identified two main themes and 8 sub-themes, which we explore in Section \ref{results}.

\subsection{Participatory Methods: Feedback  \& Roundtable}\label{participatory}

Throughout our research process, we prioritized the participation of relevant expert groups as a key element of our methodological approach: what social scientists describe as \textit{participatory design} or \textit{participatory action research}  \cite{Whitney, bodker_participatory_2022, hayes_relationship_2011, cooper_systematic_2022}. In the language of early participatory design scholar Susanne B\o{}dker, our work "takes its starting point from the values and concerns of [a] particular group[]" (here, public interest technology practitioners) and "aims for... mutual learning between designers and people" \cite[p.~6~\&~7]{bodker_participatory_2022}. As such, we shared an early draft of our interview analysis with participants to both ensure accuracy and incorporate their voices. After the interview stage, we also shaped our analysis using insight from others within our organization, as well as external public interest technologists whose work came up through the process of research and interviews. In this way, we co-created "a multi-phase process with community partners to iteratively develop" our analysis while "integrating community knowledge and practices" \cite{cooper_systematic_2022}.

The capstone of this collaborative approach was a private virtual roundtable discussion with 20 leading PIT professionals which we hosted in August 2023. All interview participants were also invited to attend the roundtable; eight of the eleven interviewees ultimately joined; the demographic makeup of the roundtable is summarized in Table \ref{tab:roundtable}. The purpose of the roundtable was twofold: first, to facilitate a discussion among PIT leaders regarding the status of public interest tech within civil society contexts, and second, to consult those same PIT leaders' expertise with regards to this interview study. As such, participants were asked to read an early draft of this paper prior to the event, and the roundtable began with a recap of our methodology and our preliminary findings around the need for more robust community-building and professionalization within PIT. The discussion included group reflections regarding the evolution, state, and future of PIT. Roundtable participants were also invited to directly share how their experiences aligned or differed from the themes our prior interviews revealed (i.e., community and professionalization), and to discuss any literatures, audiences, or perspectives that we had omitted. Finally, the roundtable closed with a brainstorming session for recommendations on how to move PIT forward.
Analysis of detailed notes from the discussion were used to deepen our understanding of the themes, and is incorporated into our discussion below.

\subsection{Limitations}

\begin{table}
\caption{Summary of roundtable participants}
  \label{tab:roundtable}
    \centering
    \begin{tabular}{llc}
    \toprule
         \multicolumn{2}{c}{Category}& Number of\\
         & & Participants\\
         \midrule
         \textit{Gender}&  Woman& 7\\
         &  Man& 12\\
         &  Non-binary& 1\\
         \midrule
         \textit{Region}&  USA& 15\\
         &  Western Europe& 4\\
         &  Other& 1\\
         \midrule
         \textit{Sector}&  Civil Society& 13\\
 & Philanthropy&3\\
 & Academia&2\\
 & For-Profit&1\\
 & Government&1\\
 \midrule
 \multicolumn{2}{c}{\textbf{Total}}&\textbf{20}\\ \bottomrule
    \end{tabular}
\end{table}

Due to the systemic and cultural factors which shape who is promoted and retained within the PIT profession - much like within computing professions \cite{pop} - our study focus on senior leaders within PIT corresponded to a less diverse pool of participants with respect to race, gender, and nationality. Though efforts were made to recruit beyond simply the North American context, the participant list is indeed primarily U.S.-centric. A future study with  distinct focus on more early- and mid-career PIT professionals might more easily recruit diverse participants.

While the authors have a diverse range of experience and expertise, the design of interviews and roundtable discussions ultimately originates from the primary perspective of a technologist. Questions relating to participants' activism, research, or other valences of PIT work were not directly addressed. Relatedly, we also acknowledge - and discussed during the roundtable - the effect of a fundamental confirmation bias on our results. We did not speak, for instance, with any practitioners who wanted to join PIT but could not, or who briefly worked within PIT and chose to leave for different work. Future research might explore such boundary cases in order to better understand the logics of belonging within PIT as a nascent discipline.

\section{Interview Findings}\label{results}
We conducted this research project with the goal of contributing to a shared story - told by practitioners themselves - about the present needs and future priorities of public interest technology as a field. 
As described above, the findings shared shared below were identified via an iterative collaborative process involving preliminary thematic analysis of interviews and discussion of those findings with experts via a roundtable.
Underscoring our impetus for the project, participants consistently noted a need for improved integration of technologists into public interest organizations. 
In addition to more situational insights we gained regarding the context of PIT, two themes in particular emerged: the importance of communities of practice, and the importance of the professional pipeline into public interest technology work.

\subsubsection{Contextualizing PIT}
Participants' experiences allowed us to fill in a sketch of what kind of projects and community existed among technologists working in civic, justice-oriented, or public interest spaces prior to the influx of philanthropic investment from the NetGain Partnership. Interviews indicate that the gap in public interest technology work in the early 2000s, which we noted in the brief PIT literature review given to participants prior to their participation, was not a gap in the work so much as a shift in the landscape of digital technology in the public eye, and a reflection of the new, contextually-specific nature of framing of this kind of work as public interest technology. Multiple participants reflected that the work they were doing in the early 2000s, retrospectively, would fall under what we now call public interest technology, but that the label or concept of a professional field wasn't as prominent.\footnote{ Examples such as the US Congress Office of Technology Assessment (which ran 1974-1995) mark even earlier efforts to institutionalize public interest technology.} As Bruce Schneier puts it, ``public-interest technology isn't one thing; it's many things. And not everyone likes the term. Maybe it's not the most accurate term for what different people do, but it's the best umbrella term that covers everyone" \cite{Bruce}.

The advent of the term public interest technology is associated with influence from public interest law - one participant expressed that ``\textit{the whole public interest lens was developed by lawyers} (P3)'' - and the work of philanthropists. This accompanied a broader cultural set of changes in ``\textit{the formulation between the relationship of technology and politics} (P6)'' in the early 2000s, as well as changes in the prominence of social media and the internet on the radars of civil society, civil rights, and advocacy organizations \cite{bennett}.\footnote{Before this point, for many public-facing organizations, ``\textit{there was a lack of awareness of the value}'' of technologists, noted P1.} Some also associated public interest framing with a shift in funders and organizations prioritizing policy-based strands of the field, especially US policy. But several participants described their professional practice in terms that prioritized global community-building, and also made it clear that ``\textit{tech policy is a sliver of the public interest technology universe} (P2).''

Agnosticism around sector and domain was a common thread across participants; our interviews suggest that public interest technology as a field revolves less around specific technologies and more around ways of thinking about and working with technology. What constitutes the public interest may be context-dependent, but as Shevin et al (2022) put it, what ties together PIT work across varying understandings of the public is a combination of values-based frameworks and prioritization of communities impacted by technologies \cite{Shevin}. Understanding PIT in this way, with values-based ways of working at the center, sheds light on the varying career pathways that public interest technologists might travel. Some start their careers in technologically focused roles, and others not - what connects PIT as a field is a values-based motivation to work on issues of the public interest.

\subsection{Community}

One of the questions we were interested in, based on our background research, was whether interviewees felt they had \textit{communities of practice}, as theorized by Etienne Wenger and Jean Lave. Several interviewees prioritised this topic and emphasized the importance of communities of practice across organizations for public technologists, leading us to highlight it as a key theme in our findings. Interviewees described what attributes characterize a successful, useful community of practice \cite{Wenger, Lave}: 

\subsubsection{Relational Spaces}
 In general, we heard a need for more community spaces and opportunities for relationship building, from informal community meet-ups and mentorship to ``\textit{tech conferences} (P8)'' where participants felt able to meet other people doing similar work, share experiences, and build points of common narrative around the work they were doing. Especially for those who might be the only technologist in their organization, this is crucial: as one participant noted, without networking and relationship building, ``\textit{it's incredibly difficult to learn and grow and share things because we are so isolated} (P7).''

\subsubsection{Collaboration Across Organizations}
Where community connections are present, public interest technologists may often feel like they have multiple communities of practice (as opposed to a singular one), noting that ``\textit{it changes depending on topic} (P4).'' The communities in question are not necessarily linked to their organization or employer. Participants highlighted that a good community of practice model would incorporate the importance of collaboration, especially ``\textit{across organizations} (P6).'' One participant noted that cross-organizational collaboration allowed for ``\textit{work on broader projects} (P11);'' another brought up that cross-organizational communities of public interest technologists were able to support each other's work where possible using ``\textit{the resources of the community} (P9).'' Several expressed that part of what made their community or communities of practice useful was the opportunity to collaborate with a group of people with diverse backgrounds: people whose interests ``\textit{don't fit the mold,} (P3)'' a ``\textit{mix of technologists and people who work at lawyers' offices and some journalists who hang out there} (P6),'' as well as \textit{``economists and policy and legal people with technology skills at the other side of the equation.} (P10)''

\subsubsection{Teaching and Learning}
Multiple participants said that an effective community of practice would prioritize the ability to teach and learn within the community, and foreground ``\textit{knowledge sharin}g (P11)'' among members. One participant described the importance of maintaining ``\textit{connections to the local grassroots communities that....created these independent learning environments where we learn and teach each other} (P5).'' The technologists we spoke to, across the board, emphasized that having opportunities to continue learning  - from having designated ``\textit{learning time} (P11)'' within working hours, to having the opportunity to learn from other technologists - was a crucial part of succeeding in their work. As one said, it has \textit{``affected and changed my life to have access...to conferences and skills to remain in this field} (P7).'' Another noted, ``\textit{I always wish I had more time to learn myself and learn from other technologists }(P5).''

\subsubsection{Common Values}
Another crucial element to a useful community of practice in public technology is the presence of common values shared among community members. A sense of a lack of community can arise when technologists understand that they are ``\textit{not doing the work for the same reasons} (P8).'' In a field where day-to-day work can directly impact urgent questions of human rights, many feel community building requires that ``\textit{you have to be invested in it from a person and community level} (P5),'' as opposed to just thinking about ``\textit{what's the profit} (P9)'' in the work \cite{values}. However, for some participants, this raises questions around building communities of practice for public interest technologists across sectors. One noted that a pathway between the nonprofit and for-profit sectors and ``\textit{between technology providers }(P4)'' was missing from the current public interest model. That gap raised questions around what a career path in public interest technology could look like. 

\subsection{Professionalization}

Relatedly, the second dominant theme from our interviews was reflection around the \textit{pipeline} into and through public interest technology work. Interviewees noted a place for pipeline development in response to key challenges and needs the field currently faces: 

\subsubsection{Legitimacy}
We noticed a strong relationship between pipeline development, recognition of public interest technology as a field, and the ability to find professional legitimacy doing public interest technology work. One participant summed up that ``\textit{we still don't have a great model for people who want to do this in their lives. What do you study? Where do you go after? We're all creating our own way forward. It makes it hard to get people attracted to the field} (P2).'' We choose to mention legitimacy here as a distinct theme, though it is clearly connected to and embedded in discussion of other themes (e.g. it seems to be a component of how interviewees understand legibility). The precise conceptual relationships between these themes could be fruitful to explore further, but for the purposes of our study we focus on what these themes tell us about the state of the practice.

\subsubsection{Legibility}
Participants cited challenges in making public interest technology work legible to stakeholders outside technologist communities, in part because ``\textit{people have no clue that this role exists} (P8).'' As a result, public interest technologists may not be seen as doing legitimate work - as one participant put it, ``\textit{they're not valorised} (P9).'' As another said, ``\textit{we're not lacking imagination. We're lacking understanding of the different lanes and what the broader spectrum of public interest is, and who is in what lane, and how that work is done more broadly} (P6).'' Addressing this gap would thus involve, among other things, prioritization of how public interest technology work is framed and communicated to stakeholders outside the field, through a concrete effort ``t\textit{o bring in more people who can span that gamut} (P11)'' between technical understanding of public interest technology work, and the ability to effectively communicate its value. 

However, existing initiatives to develop and streamline a public interest technology pipeline are seen to be moving in the right direction: it is ``\textit{easier to speak}'' as someone affiliated with an organization versus an individual ``\textit{crazy privacy advocate} (P8).'' Our interviews indicate that strengthening an effective pipeline into public interest technology work would require building skill and comfort around the variety of work types that public interest technologists do and the multiplicity of roles they may serve, often all in the context of one professional position. 

\subsubsection{Siloing}
Currently, as a means of navigating the variety of types of work that public interest technologists do, many organizations silo off technologists into a separate department or team. But - echoing indications from the existing organizational PIT research - participants advocated for future field-building work to move away from this trend, noting ``\textit{real concern that we create silos...that we recognise that there is this intersection between policy and technology decisions and we create silos to try to address the problem} (P1).''

Instead of silos, some participants expressed desire to see technologists integrated into organizations in ways that reflected the multiple areas their work impacted. This is a priority for technologists in leadership positions - for example, one participant shared that when ``\textit{I was CITO...we de-siloed a whole bunch and rearchitected the organization} (P10).'' Embracing the multiplicity of workstreams that technologists may be involved in seems key. Looking forward, ``\textit{technology understanding needs to permeate government. There needs to be a technologist on any office that works on policy touching technology and (...) there should be policy people embedded in the technology development shops and ideally there is a policy person lawyer working within every product development team} (P1).''

\subsubsection{Access}
Participants noted the importance of supporting multiple, and especially non-technical, pathways into public interest technology - this impacts equity of access to the field. Several participants hesitated to call themselves technologists, with several expressing that their formal education was not related to digital technology nor computer science, echoing a broader trend that many public interest technologists follow non-traditional, non-technical paths into public interest technology work. 

We note a sense of \textit{directionality} into public interest technology work, with multiple participants bringing up the feeling that effective public interest technology work often starts from the public interest side. One participant recalled, ``\textit{we used to have the saying: it's much easier to take someone who is a movement organizer and an activist and teach them about technology than to put a heart into an engineer} (P5).'' Others agreed, noting that for many, the path into ``\textit{policy seems harder}, (P9)'' but that ``\textit{it's not that hard}'' to introduce technical concepts to ``\textit{non-tech audiences.} (P2)''  Notably, this description of directionality was controversial for some participants in the virtual roundtable. While some felt that it accurately acknowledged how engineers being inappropriately put in charge of sociopolitical projects (e.g. immigration, refugee services, etc.) can be a serious source of harm, others found it to be inaccurate and ostracizing. Further investigation into the dynamic between engineers and activists might reveal why statements like this are such a flashpoint, and might point towards avenues to pursue deeper mutual respect across disciplines.
Finally, one participant felt that the ``\textit{biggest gaps}'' in the current public interest model are ``\textit{geographical},'' and that funding structures can exacerbate a sense of disconnection for technologists, for example those working outside the remit of ``\textit{US foreign policy} (P8).'' 

\section{Discussion}\label{discussion}

These findings show that it is imperative to define and specify public interest technology in a manner informed by technologists already charting the path into public interest work. Formal professionalization and pipeline development would serve to fill gaps around the legitimacy of PIT work, making PIT more legible to public-serving institutions and deepening the understanding and accessibility of public interest technology as a framing for work being done. The further development of communities of practice in public interest technology, especially those not necessarily linked to a specific sector, would serve to build stronger possibilities for sharing of diverse knowledges and expertises across public interest technologists, strengthening technologists' ability to form disciplinary and professional ties to other technologists in similar spaces of work and mitigating the disconnect that might exist for someone who, for example, is the only technologist in their organization. 

\subsection{Professionalization}
Interview participants emphasized how crucial \textit{pipeline development and professionalization} are in terms of the formalization of public interest technology work. As more and more public interest organizations find themselves desiring technologists, it serves the needs of those organizations (and those technologists) to develop a way to talk about, recognize, and legitimize the work that technologists are doing. Indeed, professionalization directly confronts the challenges of fit and scope identified in previous work (see \S\ref{challenges} and \citet{Chock}). Legitimizing and developing a pipeline for PIT as a field also incentivizes individuals to work in the public interest, and facilitates transitions into public interest work, not least by creating a shared and accessible understanding of the value of PIT work and the creation of (multiple\footnote{As we note elsewhere, pipeline development efforts should seek to build up and highlight the diversity of paths that can lead into PIT work, rather than impose one ``right'' way.}) clearer paths for career progression within PIT work \cite{here-to-there}. Roundtable participants agreed, too, that the absence of incentives, reward structures, and metrics of success and legitimacy was a major obstacle for individuals as they pursued PIT careers. 

An important prerequisite to professionalization is the allocation of funding. Taking as an example the Ford Foundation's pipeline-building for public interest law as a field in the 1960s-1970s, a 2016 Freedman Consulting report highlights \textit{adequate funding} as a key factor in nascent PIT initatives \cite{here-to-there}. Creating funded pathways from private to public work (for example, through pro bono programs) proved important to developing public interest law. Alleviating the financial pressures associated with non-private sector jobs - through federal loan forgiveness for recent graduates and more competitive compensation - are also considered enabling factors for the growth and professionalization of public interest work beyond the technological domain\cite{Trubek}.

Current efforts in the field are moving in the right direction. As of 2023, 64 American higher education institutions have established public interest technology initiatives as members of the Public Interest Technology University Network, spearheaded by New America, the Ford Foundation, and the Hewlett Foundation. From New York University's first PIT career fair to the opening of a Public Interest Technology Lab led by Dr. Latanya Sweeney at Harvard University, to the emergence of dedicated degree programs and courses, the past few years alone have seen universities charting out new ways for students and researchers to engage with public interest technology \cite{Andreen}. Outside of the university, funded PIT fellowships are helping technologists find their way into public interest contexts. As detailed by Mutale Nkonde, groups like Public Knowledge and the Ada Lovelace Institute have created fellowship roles that allow public interest technologists to be integrated into the organization's goals. Public Knowledge, for example, has a recurring role for a Communications Justice Fellow. In 2018, the role was filled by Alisa Valentin, who worked on broadband connectivity access policy, with particular focus on marginalized rural communities. And at the Ada Lovelace Institute, Os Keyes used their 2018 Fellow position to work on public sector technology initiatives like anti-facial recognition advocacy \cite{Mutale}. Groups like All Tech Is Human, founded in 2018, are broadening access to these emergent pipelines through programs like mentoring and job-boards \cite{ResponsibleTech}.

Some limitations we encountered in this project further highlight the need for pipeline development. Recruitment and retention challenges of public interest technologists are often based in resource scarcity and mirror those in the field of public interest law: competitive compensation and benefits and pro-tech culture, as they directly compare with the private sector \cite{Freedman2016, Trubek}. Freedman (2016a) notes an expansive list of possible \textit{professional interventions} (supply-side, demand-side, and market-focused) to assuage such challenges.
Some of the suggested changes include making job descriptions more accurate and appealing to technologists, facilitating the placement of technologists at public interest organizations (through, e.g., placements or consulting facilitation), and making changes to how technologists get contracted to enable better ``skill deployment'' at public interest institutions \cite{Freedman2016}.

Such considerations around recruitment, retention, and their consequences were very present concerns for our participants. We found that the group of public interest technologists who had been in the field long enough to now be in leadership roles - and whose work was recognized by promotion into leadership roles, and who did not leave the field due to limited opportunities for growth - was not particularly diverse. Field-building efforts at multiple levels of career seniority could start to change this. By increasing the legibility of public interest technology as a legitimate career path, pipeline development could mitigate one form of under-valorisation that public interest technologists from underrepresented communities face. Furthermore, roundtable participants expressed concern regarding the pitfalls of technical professionalization, pushing back against a normative entrance point to PIT coming from an engineering-based curriculum and emphasizing that a successful PIT pipeline elevates not only technologists whose work touches on advocacy, but also activists and those who work in the public interest whose work touches on technology.

\subsection{Community Building}
Our findings further support a need for shared understanding of public interest technology work within communities of technologists - in this way, communities of practice and more informal shared spaces for public interest technology work may help in seeing through more effective integration of public interest technologists. Community building  has the potential to address the challenges of both fostering and finding individuals with multi-disciplinary expertise in both technical and societal issues, as identified in previous work (see \S\ref{challenges} and \citet{Freedman2013}). 

The creation of clearer professional pipelines into PIT, then, should be prioritized alongside the fostering of more accessible, open, and diverse spaces for technologists to share experiences and collaborate on work in the public interest. 
Indeed, one demonstrated goal and benefit of focused pipeline creation and strengthening efforts such as university programs and fellowships is a more cohesive community. As evidenced by the fact that 62\% of interviewees in Costanza-Chock et al.'s \#MoreThanCode report cited a "supportive individual" and 18\% cited a "mentor" as important paths into PIT work, increased and intentional community-building are key components to strengthening the pipeline and the field \cite{Chock}. 

Additionally, the creation and strengthening of PIT communities that weave through and outside of sector boundaries (e.g., civil society, government, private sector, and journalism) is an important part of creating sustainability for any professionalized pipeline for PIT work - as Chock et al. discuss in their 2018 study \cite{Chock}. Public interest technologists and New America fellows Afua Bruce and Maria Filippelli also note that strategies involving \textit{mutual education between technologists and civil rights organizations} are important to strengthening mutual understanding. Other similar strategies for integrating public interest technologists into public interest organizations include ``defining common language...[and]...scaling efforts,'' the latter of which might include the creation of more opportunities for civil rights leaders and technologists to share space and research \cite{TheHill, Afua}. In our interviews, technologists emphasized how ``nontraditional'' professional backgrounds led them to success in public interest technology work - this is a strength of public interest technology as a budding field, and one we should seek to bolster rather than erase.

Positive examples of community building exist at multiple levels, from formal convenings like Ford Foundation's The Table, to grassroots working groups such as the internet standards Public Interest Technology Group, to the informal community meetups documented by Civic Hall's Civic Tech Field Guide.\footnote{See "Lessons from the Table: Civil rights, technology, and privacy," a 2019 Learning Reflections blog post by Lori McGlinchey for the Ford Foundation: \url{https://www.fordfoundation.org/work/learning/learning-reflections/lessons-from-the-table-civil-rights-technology-and-privacy/}; the Public Interest Technology Group: \url{https://pitg.network/}; and the Civic Tech Field Guide: \url{https://civictech.guide/}.} For example, initiatives like Technologists for the Public Good curate events and working committees of technologists working from shared values including ``community,'' ``collaboration,'' and ``knowledge sharing."\footnote{See "Technologists for the Public Good:" \url{https://www.publicgood.tech/}} The organization Simply Secure hosts a Slack community and holds monthly ``human rights centered design'' calls which create space for people to connect.\footnote{See "Human Rights Centered Design:" \url{https://hrcd.pubpub.org/}}. Individuals within the field also continue to create space for community, from organizing regular happy hours that provide a crucial, informal space to connect and network, to personally running PIT mailing lists and mentoring others.

Development of communities of practice could also help address patterns that limited the scope of this project - when identifying interview participants, we found widely disparate views as to whether or not public interest technologists can be located within the private sector as a matter of definition, and if so, the challenges in reaching those public interest technologists. Both of these debates underscore our questions around how communities of practice form and how membership in them develops. 

\subsection{What PIT Offers}
In addition to these areas of growth, roundtable participants emphasized the importance of acknowledging the strength of what PIT already has to offer. For instance, the discussion characterized various unique mindsets and approaches that they asserted are common across PIT: prioritizing public interest values rather than neoliberal efficiency; inclusive sensemaking and systems thinking rather than top-down problem solving; appreciating complexity rather than linear reductionism; doing ``with'' rather than doing ``for'', and prioritizing needs and perspectives of those closest to the problem. What's more, the existence of multifaceted PIT individuals helps to break down stubborn binaries between `good' and `bad' among government, tech companies, and civil society. Indeed, centering PIT individuals as a frame might help to parse out and examine the persistent tension between engineering and advocacy frames that is a microcosm of broader societal and epistemological tensions-as explored in prior work by Asad \& Le Dantec and Bennett \cite{bikes, bennett}. Finally, participants offered that the breadth encompassed within the term ``public interest,'' while a source of confusion and consternation at times, is nevertheless a strength for the flexibility it allows.

\section{Future Work and Conclusions}\label{conclusion}

Our research explores the significance of PIT, an explicitly interdisciplinary field at the nexus of technology and policy that is gaining momentum among today's sociotechnical and sociopolitical practitioners. Drawing upon interviews and discussions with public interest technology practitioners with diverse training and professional experience in civil society, we show that technologists who have the support of leadership and the ability to issue spot and to proactively set strategy can serve in their roles more effectively. Our findings also indicate the importance of both formal and informal spaces of community building. Finally, established experts in the field shared the need for more professionalization. These findings lay a foundation not only for material improvements to PIT for those civil society professionals to come, but also for further research exploring this unique and rapidly-evolving professional space.

At the nexus of community building and pipeline development is a set of questions for further consideration regarding leadership and expertise in public interest technology. First, how PIT \textit{translates across sectors} is one area for continued exploration. While this report focuses primarily on PIT in the context of civil society organizations, our interview findings highlight the importance of moving forward with intentionality when defining how technologists working in the public, private, and academic sectors may indeed be working in the public interest.
Similarly, it would be informative to \textit{expand the geographic scope} of this work, to examine how PIT is evolving in contexts beyond the United States, or \textit{expand its professional scope}, to include early- and mid-career public interest technologists.

Further, we \textit{call on individuals in organizational leadership positions} to understand the key role they play in shaping what pipeline building and developing communities of practice will look like for public interest technology as a growing field. We underscore the importance of effective leadership in civil society organizations aiming to enable and support technologists on staff. As we have outlined in this study, replicating the successful trajectories of public interest technologists like those interviewed here is not impossible - but it does require dedicated support, strategy, and resources.

Public interest tech practitioners play diverse roles: bridging technology, engineering, civil society, policy, and government. Even while the details around what, exactly, PIT is are still being actively negotiated, our research provides valuable insight into those technologists working in civil society. As PIT gains momentum on the ground, we must not ignore its growing role within the civil sector, nor overlook its implications for the frontiers of sociotechnical research.

\begin{acks}
This work is wholly supported by the Center for Democracy \& Technology (CDT), a 25-year-old 501(c)3 nonpartisan nonprofit organization working to promote democratic values by shaping technology policy and internet architecture. Furthermore, this work was only possible thanks to the dozens of PIT professionals and experts who participated in our study and provided feedback on our findings. We specifically would like to thank Alex Givens, Dhanaraj Thakur, Gabriel Nicholas,  Joe Hall, Georgia Bullen, Sarah Aoun,  Michelle Shevin, and Cedric Whitney for reviewing and shaping our analysis.
\end{acks}

\bibliographystyle{ACM-Reference-Format}
\bibliography{Zotero_cites}


\end{document}